\documentclass[10pt]{article}
\usepackage{amssymb}
\usepackage{amsmath}
\usepackage[dvipdfm]{geometry}
\usepackage[dvipdfm]{color}

\newcommand{\eit}{e^{i\theta}}
\newcommand{\dth}{\,\frac{d\theta}{2\pi}}
\newcommand{\intt}{\int_{-\pi}^{\pi}}
\newcommand{\half}{\frac{1}{2}}

\begin{document}

\title{A Counterexample to Cover's $2P$ Conjecture on Gaussian
Feedback Capacity}

\author{Young-Han Kim}
\date{\today}
\maketitle

\begin{abstract}
We provide a counterexample to Cover's conjecture that the feedback
capacity $C_\textrm{FB}$ of an additive Gaussian noise channel under
power constraint $P$ be no greater than the nonfeedback capacity $C$ of
the same channel under power constraint $2P$, i.e.,
$C_\textrm{FB}(P) \le C(2P)$.
\end{abstract}

\section{Background}
\label{sec:intro}
Consider the additive Gaussian noise channel 
\[
Y_i = X_i + Z_i,\quad i=1,2,\ldots,
\]
where the additive Gaussian noise process $\{Z_i\}_{i=1}^\infty$ is
stationary.  It is well known that feedback does not increase the
capacity by much.  For example, the following relationships hold
between the nonfeedback capacity $C(P)$ and feedback capacity
$C_\text{FB}(P)$ under the average power constraint
$P$:
\begin{align*}
C(P) &\le C_\text{FB}(P) \le 2C(P)\\
C(P) &\le C_\text{FB}(P) \le C(P) + \frac{1}{2}.
\end{align*} 
(See Cover and Pombra~\cite{Cover--Pombra1989} for rigorous
definitions of feedback and nonfeedback capacities and proofs for the
above upper bounds. Throughout this paper, the capacity is in bits and
the logarithm is to \text{base 2}.)

These upper bounds on feedback capacity were later refined by Chen and
Yanagi~\cite{Chen--Yanagi1999} as
\begin{align*}
C_\text{FB}(P) &\le \left(1+\frac{1}{\alpha}\right)C(\alpha P)\\
C_\text{FB}(P) &\le C(\alpha P) + \frac{1}{2} 
\log \left(1+\frac{1}{\alpha}\right)
\end{align*} 
for any $\alpha > 0$.  In particular, taking $\alpha = 2$, we get
\begin{align*}
C_\text{FB}(P) \le \min \left\{\left(\frac{3}{2}\right)C(2P),\; C(2P) +
\frac{1}{2} \log \left(\frac{3}{2}\right)\right\}.
\end{align*} 
In fact, Cover~\cite{Cover1987} conjectured that
\begin{align*}
C_\text{FB}(P) \le C(2P)
\end{align*}
and it has been long believed that this conjecture is true.  (See Chen
and Yanagi~\cite{Yanagi1997, Chen--Yanagi1997} for a partial
confirmation of Cover's conjecture.)

\section{A Counterexample}
Consider the stationary Gaussian noise process $\{Z_i\}_{i=1}^\infty$
with power spectral density
\begin{equation}
S_Z(\eit) = |1+\eit|^2 = 2(1+\cos\theta).\label{eq:spec}
\end{equation}
Now under the power constraint $P = 2$, it can be easily shown
\cite[Section 7.4]{Blahut1987} that the nonfeedback capacity is
achieved by the water-filling input spectrum $S_X(\eit) =
2(1-\cos\theta)$, which yields the output spectrum $S_Y(\eit) \equiv
4$ and the capacity
\[
C(2) = \intt \half \log \left(\frac{S_Y(\eit)}{S_Z(\eit)}\right) \dth = 
1 \text{ bit.}
\]

On the other hand, it can be shown \cite{Yang--Kavcic--Tatikonda2004}
that the celebrated Schalkwijk--Kailath coding
scheme~\cite{Schalkwijk--Kailath1966, Schalkwijk1966} achieves the
feedback rate $-\log x_0$, where $x_0$ is the unique positive root of
the equation
\begin{equation}
\label{eq:poly}
P x^2 = (1-x^2)(1-x)^2 = (1+x)(1-x)^3.
\end{equation}
under the power constraint $P$.  (For the details on the performance
analysis of the Schalkwijk--Kailath coding scheme for the given noise
spectrum \eqref{eq:spec}, refer to \cite[Section 4]{Kim2004}.)

Now for $P=1$, we can readily check that the unique positive root
$x_0$ of \eqref{eq:poly} should be less than $1/2$, since $f(x) := x^2
- (1+x)(1-x)^3$ is strictly increasing and continuous on $[0,1]$ with
$f(0) = -1$ and $f({1}/{2}) = 1/16$.  Therefore,
\[
C_\text{FB}(1) \ge -\log x_0 > 1 = C(2).
\]  

\bibliographystyle{IEEEtran}
\bibliography{IEEEabrv,../../bibdb/it,../../bibdb/yhk}
\end{document}